\documentclass[aps,twocolumn,A4paper]{revtex4}
\usepackage{epsfig}
\usepackage{graphicx}
\usepackage{amsmath,amssymb,color}
\usepackage[english]{babel}

\parskip=\medskipamount

\newcommand{\eq}[1]{(\ref{#1})}
\newcommand{\fig}[1]{Fig.\ref{#1}}

\newcommand{\beq}{\begin{equation}}
\newcommand{\eeq}{\end{equation}}

\newcommand{\la}{\left<}
\newcommand{\ra}{\right>}

\begin{document}

\title{Spontaneous Symmetry Breaking and Phase Coexistence in Two-Color Networks}

\author{V. Avetisov$^{1,2}$, A. Gorsky$^{3,4}$, S. Nechaev$^{5,6}$, O. Valba$^{1,2}$}

\affiliation{$^1$N.N. Semenov Institute of Chemical Physics of the Russian Academy of Sciences,
119991, Moscow, Russia, \\ $^2$Department of Applied Mathematics, National Research University
Higher School of Economics, 101000, Moscow, Russia, \\ $^3$Institute of Information Transmission
Problems of the Russian Academy of Sciences, Moscow, Russia, \\ $^4$Moscow Institute of Physics and
Technology, Dolgoprudny 141700, Russia \\ $^5$ Universit\'e Paris--Sud/CNRS, LPTMS, UMR8626, B\^at.
100, 91405 Orsay, France, \\ $^6$P.N. Lebedev Physical Institute of the Russian Academy of
Sciences, 119991, Moscow, Russia}

\begin{abstract}
We have considered an equilibrium ensemble of large Erd\H{o}s-Renyi topological random networks
with fixed vertex degree and two types of vertices, black and white, prepared randomly with the
bond connection probability, $p$. The network energy is a sum of all unicolor triples (either black
or white), weighted with chemical potential of triples, $\mu$. Minimizing the system energy, we see
for some positive $\mu$ formation of two predominantly unicolor clusters, linked by a "string" of
$N_{bw}$ black-white bonds. We have demonstrated that the system exhibits critical behavior
manifested in emergence of a wide plateau on the $N_{bw}(\mu)$-curve, which is relevant to a
spinodal decomposition in 1st order phase transitions. In terms of a string theory, the plateau
formation can be interpreted as an entanglement between baby-universes in 2D gravity. We have
conjectured that observed classical phenomenon can be considered as a toy model for the chiral
condensate formation in quantum chromodynamics.
\end{abstract}

\keywords{topological order, Erd\H{o}s-Renyi network, phase transition, metastability, quantum
gravity, baby-universe}

\maketitle

\section{Introduction}

During last two decades it has been recognized that Landau classification of phase transitions does
not cover all patterns of symmetry breaking. The fractional quantum Hall effect and the Levin-Wen
string networks \cite{levin-wen} provide well known examples of topological order for so-called
"gapped states". Corresponding quantum phase transitions exist at $T=0$ and do not have any local
order parameters. Some symmetry aspects of 2D topological phases with $\mathbb{Z}_2$ symmetry have
been discussed in \cite{kapustin}. Quantities describing topological phases are: i) the degeneracy
of the ground state (which depends on the topology of the system), and ii) the holonomy of the
non-Abelian Berry connection. Another way to identify "topologically ordered" states is to use the
topological entanglement entropy \cite{kitaev}. In any cases, the topological order is manifested
in emergence of long-range correlations in the system. The interplay between some "local rules" and
"global constraints", influence the behavior of the whole system and is responsible for the
long-range order.

Colored random networks have become an ubiquitous new paradigm for wide range of physical and
social phenomena in distributed systems, spread from producers-consumers relations to string
theory, and from budding in lipid membranes to baby-universes creation in cosmology. Here we study
random black-white vertex networks with Levin-Wen type of the Hamiltonian \cite{levin-wen},
evolving by bond reconnections and tending to increase the number of unicolor triples under vertex
degree conservation. The "local rule" is the conservation of the vertex degree in all nodes of the
network, while the "global constraint" is the attempt to maximize the number of triples of the same
color. As the energy of triples increases, the network breaks into two predominantly unicolor
(black and white) clusters, connected by a bunch of links between black and white vertices.
Unexpectedly, this bunch is stable in a wide interval of triples energy. We conjecture that this
novel statistical phenomenon has origin relevant to quantum entanglement.

Ensembles of random Erd\H{o}s-Renyi topological graphs (networks) provide an efficient laboratory
for testing various collective phenomena in statistical physics of complex systems, being also
tightly linked to random matrix theory. Besides investigating "typical" statistical properties of
networks, like vertex degree distribution, clustering coefficients, "small world" structure and
spectra of adjacency matrices, last two decades have been marked by rapidly growing interest in
more refined graph characteristics, such, as for example, distribution of triadic motifs in
oriented networks (small subgraphs involving different triads of vertices).

Triadic interactions, being the simplest interactions beyond the free-field theory, play crucial
role in network statistics. Apparently, just presence of triadic interactions is responsible for
emergence of phase transitions in complex distributed systems. First signature of a phase
transition in a random network, known as Strauss clustering model \cite{strauss}, has been treated
by RMT in \cite{burda} and identified with the 1st order phase transition in frameworks of
mean-field cavity-like approach in \cite{newman}. Another example of phase transition is connected
with the triadic motifs pattern formation, known as "motifs superfamilies", in real evolutionary
networks \cite{alon}. The last problem has been theoretically analyzed in \cite{valba}, where it
was conjectured that stable motif profiles constituting superfamilies \cite{alon} may correspond to
stability islands associated with localized states in the space of motifs. The localization of
states occurs as a 1st order transition if the chemical potential associated with triadic motifs
energy is large enough. The critical behavior of triadic motifs concentration, as a function of the
chemical potential, has been studied in details in \cite{valba}.

Specifically, in this work we investigate the dependence of an equilibrium number of links,
$N_{bw}$, connecting "black" and "white" vertices in networks with fixed vertex degree, $p$, as the
function of the chemical potential, $\mu$, controlling the number of unicolor vertex triples (all
three vertices are black, or all three are white). The attempt to minimize the energy of unicolor
triples leads to almost immediate color separation, when at any infinitesimal positive value of
$\mu$ two clusters with opposite colors ("$\mathbb{Z}_2$-charges") are formed. More striking
phenomenon is observed for large enough $\mu$: the $N_{bw}(\mu)$-curve has a wide plateau
separating two Arrenius-type dependencies at low and high $\mu$. Such a plateau is never seen (for
any $p$) if the energy of the system is the sum of unicolor vertex pairs (but not triples), or if
the vertex degree is not conserved. We suggest two parallel interpretation of the phenomenon
observed in our simulations: "classical" (in terms of classical 1st order transitions), and
"quantum" (in terms of spins interacting with 2D topological gravity). Describing the system in
terms of classical statistical physics, we interpret the plateau emergence as the spinodal
decomposition similar to the vapor-liquid first order phase transition.

On the other hand, the "quantum gravity language" implies the realization of the network as the
triangulation of 2D fluctuating surface. Since the network does not involve metric, one deals with
purely topological version of 2D gravity where the chemical potential of unicolor triples, $\mu$,
can be understood as a 2D cosmological constant. The spin (i.e. color) separation phenomenon is
interpreted as an anomalous transport on the string worldsheet due to the axial anomaly in the
gravitational field. In terms of 2D gravity, it is conjectured that the plateau formation occurs as
the specific "entanglement" of two baby universes. We propose the synchronization mechanism of
phase transitions in each universe and suggest that their entanglement results in a sort of a phase
coexistence phenomenon. To describe the phase coexistence properly, one could try to find gravity
counterparts of thermodynamic variables. This is a subtle issue and we rely on recent
interpretation of $(P,V)$ variables in cosmological terms \cite{dolan} deling with Van-der-Waals
thermodynamics of charged or rotating black holes \cite{vdv}.

We also mention useful analogy with the chiral symmetry breaking in the instanton--antiinstanton
ensemble in quantum chromodynamics (QCD) described by random matrix model. In this case case, the
system at small $\mu$ and low temperature is in the phase with the spontaneously broken chiral
symmetry characterized by the nonvanishing chiral condensate, $\la \bar{\Psi}\Psi \ra$. Staying at
$T=0$ and increasing the chemical potential, $\mu$, we force the system to undergo a quantum phase
transition at some $\mu_c$. The density of chiral condensate is proportional to the number of
instanton--antiinstanton connections, and the chiral symmetry restoration signifies emergence of
another condensate, $\la \Psi\Psi \ra$. We conjecture that our two-color network could be a toy
model for the formation of such condensate.

\section{The model and key observations}

The setup of the model is as follows. Take $N$ vertices and label them by integers $1,...,N$.
Vertices $i=1,...,cN$ ($0<c<1$) are "black", i.e. are associated with Ising spins $\sigma_i=-1$,
while vertices $i=cN+1,...,N$ are "white" and carry $\sigma_i=+1$. The initial realization of
Erd\H{o}s-Renyi network is prepared by connecting any randomly taken pair of vertices with the
probability $p$, ($p>\frac{1}{N}$, i.e. $p$ is above the percolation threshold). Then, one randomly
chooses two arbitrary links, say, between vertices $A$ and $B$ ($A$--$B$) and between $C$ and $D$
($C$--$D$), and reconnect them, getting new links $A$--$C$ and $B$--$D$. Such reconnection
conserves the vertex degree \cite{maslov}. Now one applies the standard Metropolis algorithm with
the following rules: i) if after the reconnection the number of unicolor vertex triples is
increased, a move is accepted, ii) if the number of unicolor vertex triples is decreased by $\Delta
N_{\rm tripl}$, or remains unchanged, a move is accepted with the probability $e^{-\mu \Delta
N_{\rm tripl}}$. We have run the Metropolis algorithm repeatedly for large set of randomly chosen
pairs of links, until it converges. For one-color networks it was proven \cite{reconnection} that
such Metropolis algorithm converges to the Gibbs measure $e^{\mu N_{\rm tripl}}$ in the equilibrium
ensemble of random undirected Erd\H{o}s-Renyi networks with fixed vertex degree. For two-color
networks the convergence is not proven rigorously, however extensive numeric tests do not show any
pathologies.

\begin{figure}[ht]
\centerline{\includegraphics[width=8cm]{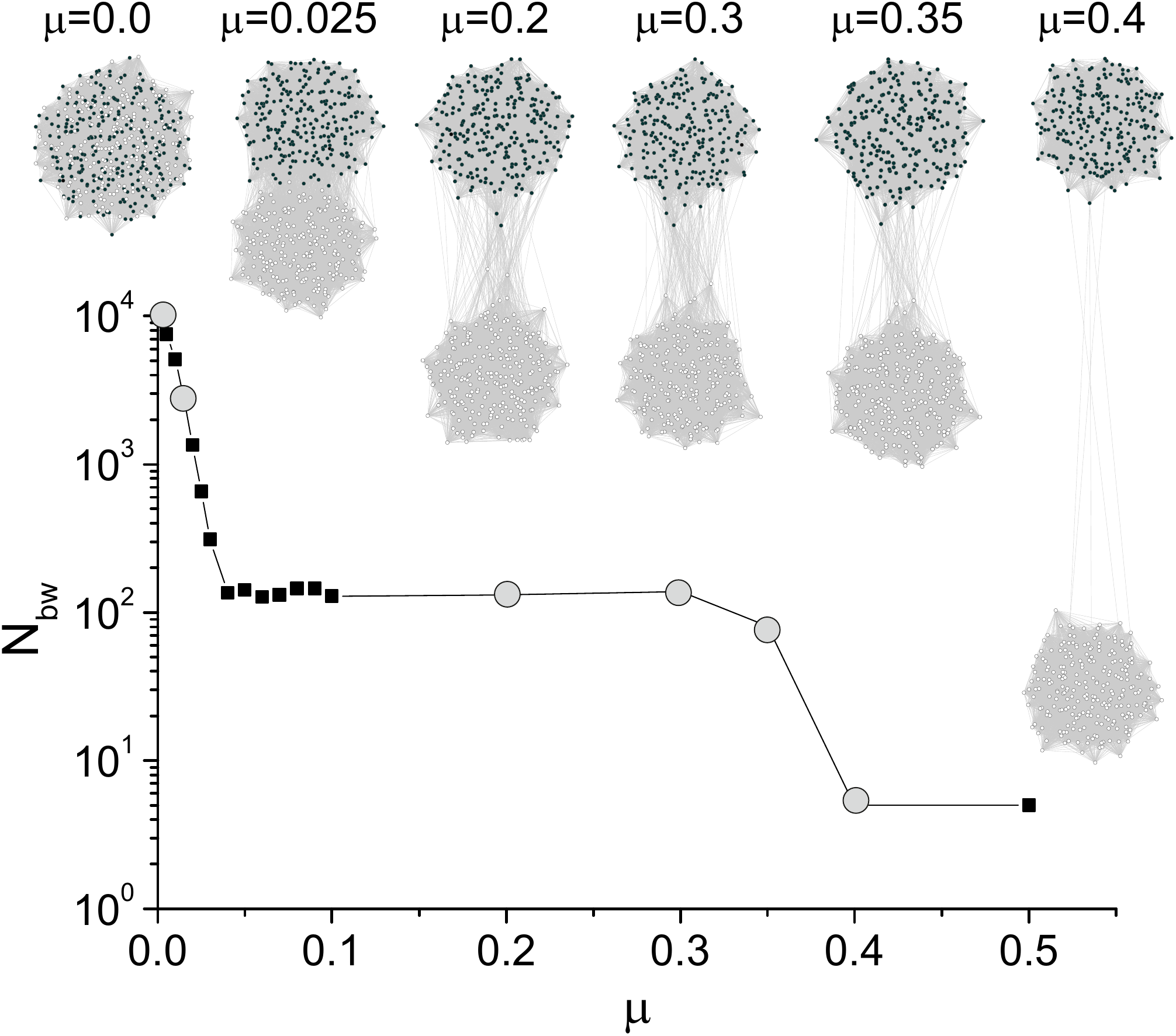}}
\caption{Dependence of equilibrium number $N_{bw}$ of black-white bonds on the chemical potential,
$\mu$, per each unicolor triple for $N=500$-vertex random network of the probability $p=0.15$ with
equal composition of black and white nodes; the typical network topologies for some values of $\mu$
along a $N_{bw}(\mu)$-curve are shown above the $N_{bw}(\mu)$-curve.}
\label{fig:01}
\end{figure}

The obtained results are as follows. The dependence of number of links connecting black and white
vertices, $N_{bw}(\mu)$, demonstrates for small and large $\mu$ the Arrenius activation kinetics,
$N_{bw}(\mu) \sim e^{-\beta(p) \mu}$, where $\beta(p)$ is some connectivity-dependent constant.
However, the function $N_{bw}(\mu)$ develops a plateau in a wide region of $\mu$, manifesting a
kind of a critical behavior. The corresponding plot for the network of $N=500$ vertices is shown in
the \fig{fig:01} for $p=0.15$. Analyzing the clustering structure of the network for different
$\mu$, one sees that at small $\mu$ ($\mu\sim \frac{1}{N}$), the system splits for any $p\gg
\frac{1}{N}$ into two clusters of predominantly black and white vertices, meaning the spontaneous
$\mathbb{Z}_2$-symmetry breaking (the clusters are allocated by preferential number of links
entering "inside" it, rather than going "outside"). So, the plateau is developed in the
$\mathbb{Z}_2$-broken phase. These results are reproducible for various values of $c$ (where $c$ is
the fraction of black vertices). The same behavior is seen for random regular networks, which have
fixed vertex degree, $d$, in all network nodes.

Remind that we consider very dense networks, $p=O(1)$, far above the percolation threshold,
$p=\frac{1}{N}$. Certainly, even in this regime the entire network contains some number of disjoint
clusters, exponentially suppressed in their sizes. In colored network, weights of different
rewirings are color-dependent and could affect the distribution in cluster sizes. We believe that
for dense networks (considered here) effect of color-dependent re-distribution in cluster sizes is
negligible against the background of exponentially suppressed overall number of disjoint
components. However, near the percolation threshold this effect might be visible and definitely
deserves special attention.

The following two key result are the subjects of our discussion and interpretation in terms of
lattice gases and string theory:
\begin{itemize}
\item Immediate spontaneous formation of black and white clusters (breaking
$\mathbb{Z}_2$-symmetry) at $\mu\sim \frac{1}{N}$;
\item  Emergence of a wide plateau in $N_{bw}(\mu)$-dependence
\end{itemize}

\section{Symmetry breaking}

\subsection{Statistical interpretation}

First of all, it is worth pointing out that the $\mathbb{Z}_2$--symmetry breaking and the plateau
formation are different physical effects. Separation of clusters is typical phenomenon which can be
understood using naive mean-field arguments. Let $N_b$ and $N_w$ be the number of black and white
nodes ($N_b+N_w=N$). In a mixed system the interaction energy of one $b$-node with $m$ randomly
chosen other nodes is $u_b=\frac{m}{N}(N_b \phi_{bb} + N_w \phi_{bw})$, where $\phi_{bb}$ and
$\phi_{bw}$ are the $b$--$b$ and $b$--$w$ interaction energies. Respectively, $u_b=\frac{m}{N}(N_b
\phi_{bb} + N_2 \phi_{wb})$. The free energy of the system reads $F=\frac{1}{2}(N_b u_b + N_w u_w)
- k T \left(N_b \ln \frac{N_b}{N} + N_w \ln \frac{N_w}{N} \right)$. Defining $\phi=\phi_{bb}+
\phi_{bw}$ and $\Delta=\phi_{bb}-\phi_{bw}$, passing to concentrations, $c_{b,w}=\frac{N_{b,w}}{N}$
($c_b+c_w=1$), and introducing the asymmetry of the network composition, $\eta=c_b-c_w$, we can
write the network free energy, $f=\frac{F}{N}$, as follows:
\begin{multline}
f(\eta) = f_0+\frac{m \Delta}{4}\eta^2 - \frac{kT}{2}\big[(1-\eta)\ln(1-\eta) \\ + (1+\eta)\ln
(1+\eta)\big]
\label{eq:01}
\end{multline}
where $f_0=\frac{m}{4}\phi + kT\ln 2$.

The system described by the thermodynamic potential \eq{eq:01}, experiences the standard phase
transition with the stationary states, $\eta_c$, determined by the equations $\frac{d
f(\eta)}{d\eta}=0$ and $\left.\frac{df(\eta)}{d\eta}\right|_{\eta_{c}}>0$. Supposing that
$|\eta|\ll 1$, we can expand $f(\eta)$ up to the 4th term and get the equation
\beq
\left(\frac{m \Delta}{2}-kT\right)\eta-\frac{kT}{3}\eta^3=0
\label{eq:02}
\eeq
having a single solution $\eta_c=0$ at $m\le \frac{kT}{\Delta}$ and exhibiting spontaneous symmetry
breaking into two phases $\eta_c^{a,b}=\pm \sqrt{3\left(\frac{m\Delta}{kT}-1 \right)}$ at $m>
\frac{kT}{\Delta}$.

The mean-field arguments for accounting a plateau, are provided in the next section in terms of
lattice gas statistics, random matrix and string theories. Being applied to networks, the random
matrix approach means the discretization of a two-dimensional string worldsheet embedded in some
D-dimensional target space. Since the colored (black and white) vertices can be described by Ising
spin variables, our system is some specific model of a matter (Ising spins) coupled to the gravity
(fluctuating surface over ensemble of allowed reconnections in the random network). In the matrix
model framework, which parallels the mean-field analysis, such kind of systems has been considered
in \cite{kazakov}, however with some important difference: in our system Ising spins are quenched,
while in matrix approaches they are typically annealed.

\subsection{Stringy interpretation}

Nowadays the string theory plays two different roles in the fundamental physics. First of all, it
is a model of quantum gravity since the spectrum of closed strings involves massless graviton.
Secondly, the string can be regarded as the "probe", and the physics on a 2D string worldsheet
should reflect all phenomena happen in entire target space, in which this worldsheet is embedded.
This is a key idea behind the gauge/string duality. Having in mind these roles of string theory and
considering a network as a discretized string worldsheet, we rephrase in stringy terms the
phenomenon observed in the black-white network.

The spontaneous symmetry breaking and the black-white cluster separation can be interpreted as
occurrence of the anomalous transport on the string worldsheet. The anomalous transport emerges due
to the quantum nonconservation of classically conserved currents in the external gauge or
gravitational fields. The spin separation phenomenon in our problem means that non-vanishing axial
current takes place at the worldsheet when the spins in sub-ensembles ${\cal S}=\{sigma=+1\}$ and
${\cal S}=\{\sigma=-1\}$ move in opposite directions. Since spins are quenched and do not
fluctuate, the current is induced purely by fluctuations of the 2D surface. The network is a
topological object, hence it is natural to describe it as the topological 2D gravity of
Jackiw-Teitelboim type  which involves the cosmological constant $\Lambda$
\beq
L= \int d^2 x \sqrt{g} \Phi( R + \Lambda),
\eeq
where $\Phi$ is the scalar dilaton field and $R$ is the Ricci curvature of the two-dimensional
metric $g$. This Lagrangian can be written as the topological 2D Yang-Mills theory with $SL(2,R)$
gauge group (see \cite{kummer} for review).

It is known that in the pure 2D gravity the total number of triangles measures the total area,
hence the corresponding chemical potential, $\mu$, of triples can be interpreted as an effective 2D
cosmological constant. At the equation of motion we have $R=-\Lambda$. There is an axial current
due to the chiral anomaly in an external gravitational field
\beq
\partial_{\nu} J_{\nu}^5 = R
\eeq
which for the equation of motion equals to the cosmological constant, $\mu$. Thus, for the axial
current itself in terms of the spin-connection $\omega_{\nu}$ we have
\beq
J_{\alpha}^{5}= \epsilon_{\alpha\nu} \omega_{\nu}  \propto \mu
\label{ax2}
\eeq
Eq.\eq{ax2} demonstrates the emergence of spin current for any nonzero chemical potential of
triads, $\mu$, which is the microscopic  mechanism of black and white nodes separation.

The $\mathbb{Z}_2$ spontaneous symmetry breaking in colored dynamic network we describe using
mean-field statistical arguments, as well as from more involved viewpoint, in terms of fluctuating
surfaces. In what follows we pass along the edge of these two different approaches.

It should be pointed out that spin separation phenomenon is known in other systems as well. At low
temperature the 2D Ising model exhibits spontaneous magnetization, developing the coherent spin
domains separated by the domain walls. The $\mathbb{Z}_2$ symmetry is spontaneously broken in this
phase. In the case of fluctuating Ising spins coupled to 2D gravity, the spontaneous creation of
"baby universes" takes place, and it is known that the spin direction in the daughter universe is
always opposite to the spin directions in the parent one (see \cite{ambjorn} for review). In our
case we see the formation of clusters with opposite spin signs due to dynamic rearrangement of
bonds only (since the spins are quenched). Similar spin separation phenomenon occurs in 2D matter
with non-vanishing chemical potential \cite{alekseev} and in 4D matter in an external magnetic
field (see \cite{kharzeev} for review). The chirality separation emerges in some lattice QCD
systems when the sheets of opposite chiralities become connected by stringy "skeletons"
\cite{horvath}.

\section{Plateau formation}

The most intriguing question concerns the interpretation of the plateau formation, since it seems
to be a phenomenon with the signature of a quantum phase transition. In contrast to the symmetry
breaking, the plateau formation is an essentially collective effect which disappears for networks
with nonconserved vertex degree.

The relevant image, which highlights the basic features of this phenomenon, is the famous puzzle
"the game of fifteen". The analogy goes as follows. When the concentration of graph links is small,
it is always possible to minimize the system energy (the sum of unicolor triples) almost locally.
This resembles the initial stages of ordering in the puzzle, when one takes care of single-dice
placements only. However, our network gets highly frustrated due to the vertex conservation
condition: in order to find the way to place some new "good" link which minimizes the unicolor
triple energy, it might be necessary to pass over high potential barrier and remove other "good"
links which have been already placed before. Thus, the energy minimization involves unfavorable
sequences of link permutations until all necessary constraints are satisfied. This is exactly what
happens on late stages of ordering in the "game of fifteen".

\subsection{Statistical interpretation}

To proceed with combinatorial interpretation, represent the graph by the symmetric adjacency matrix
$J_{ij}$ and split it in four sectors with corresponding numbers of particles (links)
$N_{bb},N_{bw},N_{wb},N_{ww}$:

\begin{tabular}{ll}\hline\hline
$N_{bb}$ & in $[\bullet\bullet]$ for $[1\le \{i,j\}\le Nc]$ \\ \hline
$N_{ww}$ & in $[\circ\circ]$ for $[Nc+1\le \{i,j\}\le N]$ \\
\hline $N_{bw}$ & in $[\bullet\circ]$ for $[Nc+1\le i\le N, 1\le j\le Nc]$ \\ \hline $N_{wb}$ & in
$[\circ\bullet]$ for $[1\le i\le Nc, Nc+1\le j\le N]$ \\ \hline\hline
\end{tabular}

\begin{figure}[ht]
\centerline{\includegraphics[width=8cm]{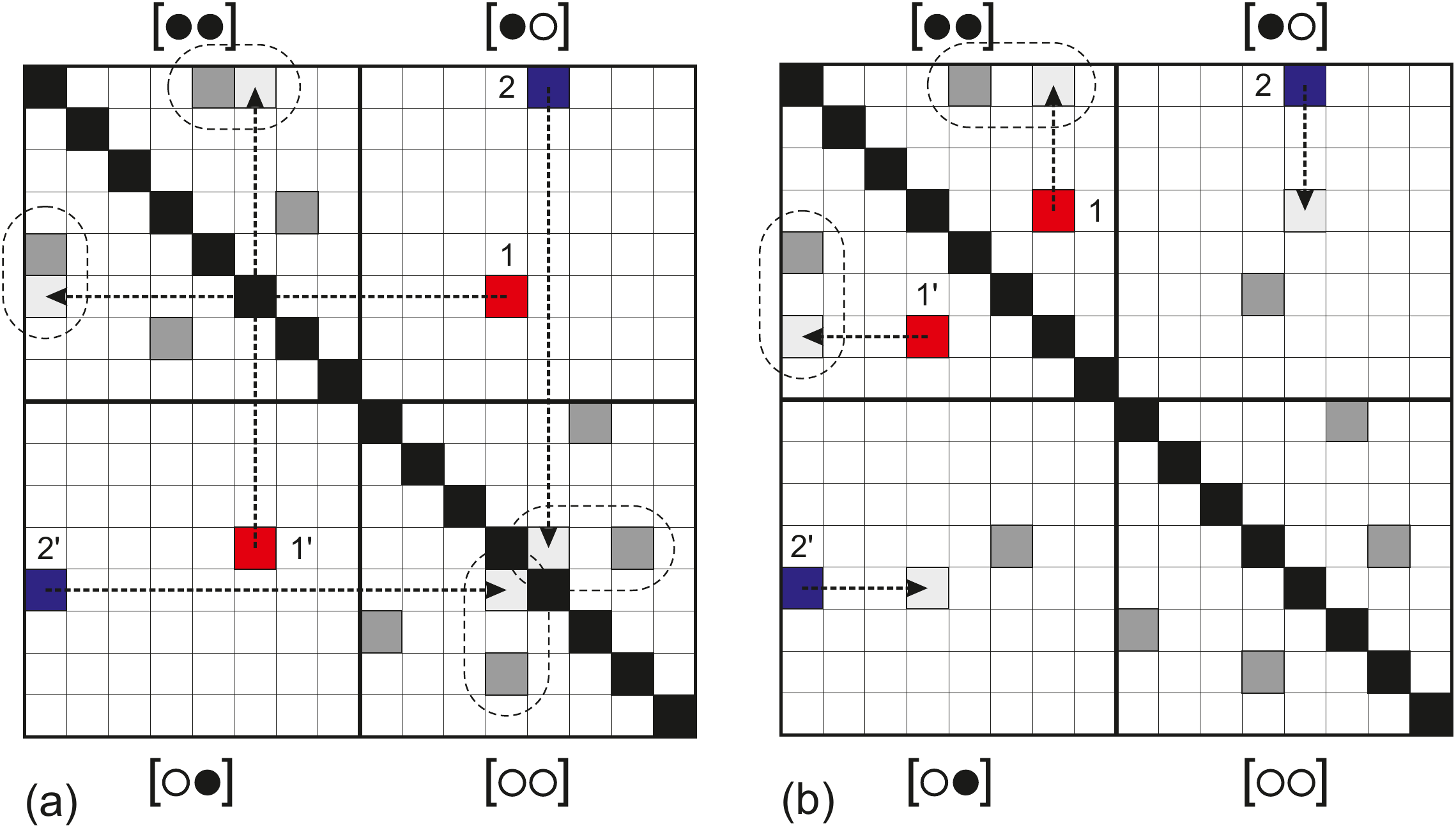}}
\caption{(a) Pulling particles (bonds) 1 and 2 (and their symmetric counterparts $1'$ and $2'$) from
the sectors $[\bullet\circ]$ and $[\circ\bullet]$ into the sectors $[\bullet\bullet]$ and
$[\circ\circ]$. The rearrangement results in two newborn triples of graph vertices (pairs of
particles); (b) Bond rearrangement which conserves number of particles in all sectors, however
increases the number of triples in the sector $[\bullet\bullet]$.}
\label{fig:02}
\end{figure}

The matrix $J_{ij}$ is symmetric, namely $N_{bw}=N_{wb}$. The unicolor triple is any pair of
particles in one row or in one column in the sectors $[\bullet\bullet]$ or $[\circ\circ]$. The
vertex degree conservation implies the condition $\sum_{i=1}^N J_{ij}=J_j^{(0)}$, where $J_j^{(0)}$
($j=1,...,N$) are the initial vertex degrees, fixed by the network preparation conditions. Hence
the simplest rearrangements of graph links should involve pairs of particles (links) shown in the
\fig{fig:02}. As we shall see, this figure helps to understand the appearance of the plateau on the
$N_{bw}(\mu)$--curve depicted in the \fig{fig:01}.

The physics behind the $N_{bw}(\mu)$--dependence is as follows. When the chemical potential, $\mu$,
of unicolor triples is increasing, the formation of aligned pairs of particles (in rows or in
columns) in the sectors $[\bullet \bullet]$ and $[\circ \circ]$ becomes favorable. There are two
mechanisms which increase the number of aligned pairs in these sectors, they are schematically
shown in the Figs.\ref{fig:02}a,b

\begin{enumerate}

\item In the first (I) mechanism the exchange of particles (shown by arrows in the \fig{fig:02}a) $1
\leftrightarrows 2$ (and of their symmetric counterparts, $1' \leftrightarrows 2'$) pushes
particles $1, 1'$ to $[\bullet \bullet]$ and particles $2, 2'$ -- to $[\circ \circ]$

\item In the second mechanism (II) -- see the \fig{fig:02}b -- the exchange $1 \leftrightarrows 2$
(and $1' \leftrightarrows 2'$) preserves the number of particles in all sectors.

\end{enumerate}

At small $\mu$ (at the beginning of the $N_{bw}(\mu)$-curve) both mechanisms (I and II) of unicolor
triples energy increasing are available. However, as the number of particles in the sectors
$[\bullet \bullet]$ and $[\circ \circ]$ exceeds some critical value, the entropic loss due
accumulation of particle in these sectors forces the 1st order vapor-liquid-like phase transition
in the sectors $[\bullet \bullet]$ and $[\circ \circ]$. Note that since the initial concentrations,
$c_{bb}^{(0)}=N_{bb}^{(0)}/(N^2c^2)$ and $c_{ww}^{(0)}=N_{ww}^{(0)}/(N^2(1-c)^2)$, of particles in
sectors $[\bullet \bullet]$ and $[\circ \circ]$ do not coincide in general (they are fixed by a
random network preparation), the difference
\beq
\left|c_{bb}^{(0)}-c_{ww}^{(0)}\right|\sim (Np)^{-1/2}
\label{eq:c0}
\eeq
desynchronizes the vapor-liquid transitions in $[\bullet \bullet]$ and $[\circ \circ]$. So, the
beginning of the plateau on the $N_{bw}(\mu)$--curve we connect with the \emph{first occurrence} of
independent vapor-liquid transitions in $[\bullet \bullet]$ or in $[\circ \circ]$. The plateau
itself we identify with the synchronization of these two vapor-liquid transitions in $[\bullet
\bullet]$ and in $[\circ \circ]$. At the plateau the mechanism I gets suppressed, because capturing
of particles by sectors $[\bullet \bullet]$ and $[\circ \circ]$ becomes entropically unfavorable.

\begin{figure}
\centerline{\includegraphics[width=8cm]{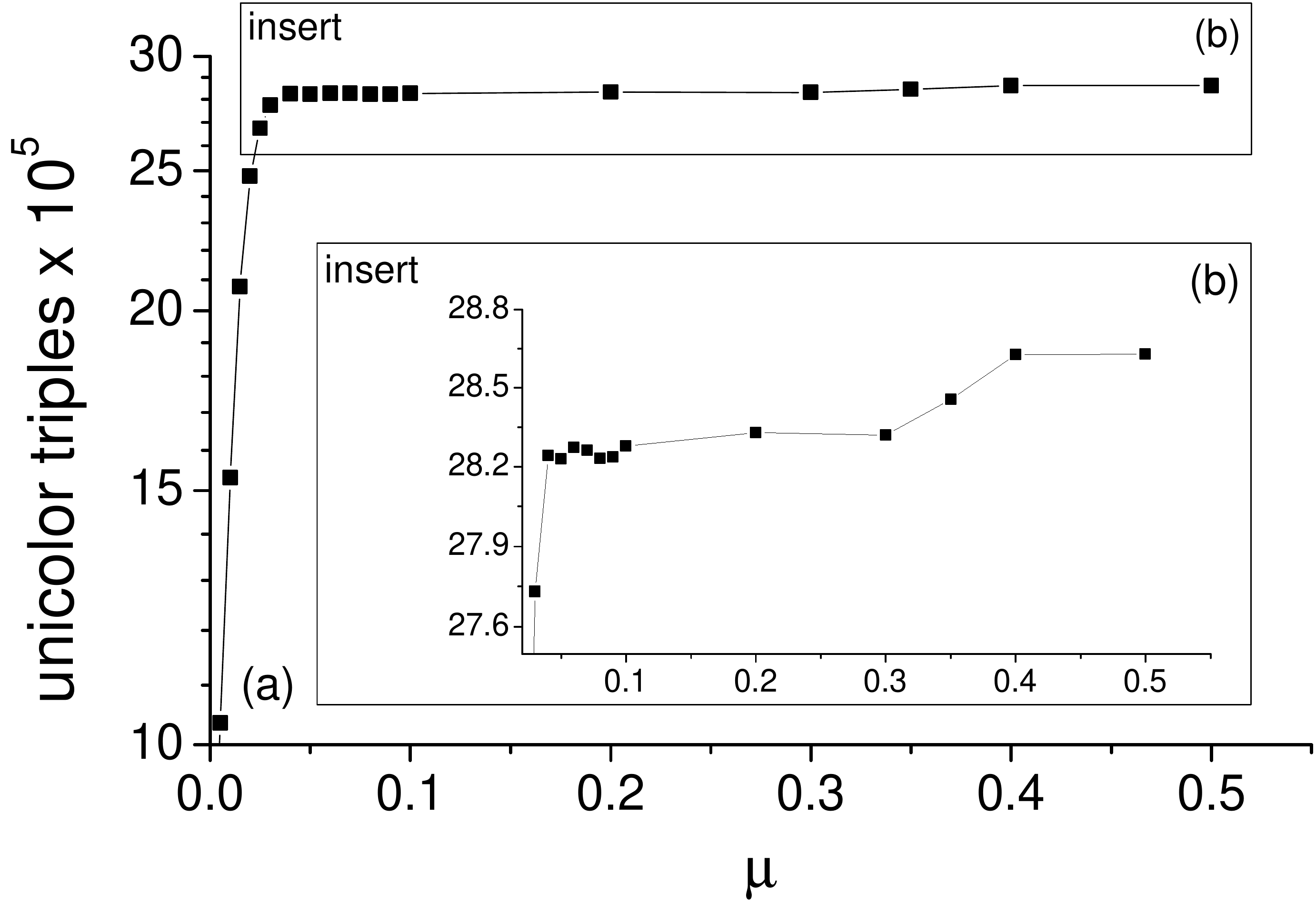}}
\caption{Number of unicolor triples, $N_{\rm tripl}$ as a function of $\mu$ in the ensemble of
networks with $N=500$ and $p=0.15$.}
\label{fig:03}
\end{figure}

The plateau on the $N_{bw}(\mu)$-curve occurs simultaneously, as shown in the \fig{fig:03}, with
the plateau on the curve $N_{\rm tripl}(\mu)$, where $N_{\rm tripl}$ is the number of unicolor
triples. The plateau on the $N_{\rm tripl}(\mu)$-curve means the topological quench of each
unicolor subnetwork in $[\bullet \bullet]$ and in $[\circ \circ]$. This behavior is a signature of
a spinodal decomposition typical for the vapor-liquid phase transitions. The mechanism which
competes with increase of energy of triples in the plateau region is just the mechanism II of
particle exchange: particle pairings, as shown in the \fig{fig:02}b, do not change the numbers of
particles in all sectors, however act against the entropy in the sectors $[\bullet\circ]$ and
$[\circ\bullet]$ since trapping of particles like 2 and $2'$ is entropically unfavorable.

Collecting all together, we are led to the following conjecture. The plateau begins at the value
$\mu_{beg}$ independently in subnetworks $[\bullet \bullet]$ and $[\circ \circ]$ as the individual
vapor-gas phase transitions. At the plateau the interaction between the sectors $[\bullet \bullet]$
and $[\circ \circ]$ is ensured by the process II, particle trapping in $[\bullet \bullet]$ and
$[\circ \circ]$ works against the entropy in $[\bullet\circ]$ and $[\circ\bullet]$, and we deal
with two effectively coupled subsystems $[\bullet \bullet]$ and $[\circ \circ]$, both exhibiting 1s
order transition and living in a joined metastable state. The plateau is finished at the value
$\mu_{end}$, at which both sectors $[\bullet \bullet]$ and $[\circ \circ]$ fall down in liquid
states, and decouple. At $\mu>\mu_{end}$ the entropic contributions in all sectors become
negligible and the Arrenius-type activation kinetics gets restored.

Note, that just above the percolation threshold, i.e. for $p\gtrsim\frac{1}{N}$ the standard
Erd\H{o}s-R\'enyi network, besides the giant component, contains many disconnected clusters with
the size distribution $c(k)\sim k^{-5/2}$, where $c(k)$ is the concentration of clusters of $k$
vertices. For black-white networks the distribution $c(k)$ should be modified since weights of
particular clusters become color-dependent. For all $p$ clusters contain only nodes of the same
color, and all black-white links are external, connecting oppositely colored clusters. The very
existence of the plateau is insensitive to the number of disjoint unicolored clusters in the
network, however the plateau disappears for quite small $p$.

\subsection{2D gravity/stringy interpretation}

In terms of 2D gravity, the emergence of the plateau can be linked with a "baby universe"
formation. It has been proven in \cite{jain,ambjorn} that critical phenomena in 2D gravity deal
with the formation of a newborn ("daughter") bubble connected by the "neck" to the parent universe.
The free energy of the system consists of two competing contributions: the mean value of the total
"neck diameter", and the spontaneous curvature of the newborn (baby) universe of a given size. The
same mechanism lies behind the "budding of vesicles" in liquid membranes \cite{lipowsky}. In terms
of the black-white network, the color separation can be interpreted as the baby universe formation,
whose size depends on the ratio $N_{bb}^{(0)}/N_{ww}^{(0)}$, where $N_{bb}^{(0)}$ (or
$N_{ww}^{(0)}$) are the initial numbers of black-black (or white-white) nodes, while the
interaction of triples plays the role of a curvature contribution to the free energy (fixed by the
cosmological constant, $\mu$).

The plateau formation, in terms of 2D gravity, signals that the neck size (or sum of all neck sizes
for multi-cluster network) connecting the parent and baby universes, does not depend in some
interval on the bulk cosmological constant, $\mu$. The pure 2D gravity undergoes the phase
transition at some value $\mu_{c}$ (see \cite{gins} for review), therefore upon the color
separation, we have two copies of 2D gravities connected by the neck. Each gravity could be in one
of two phases. Since we consider the quantum geometry, there is no reason for initial
synchronization of phase transitions in these two universes. The value of the cosmological
constant, at which the first transition happens, is presumably the point at which the plateau gets
started. Two universes are entangled, which in terms of each 2D gravity can be interpreted as the
insertion of the macroscopic loop operator.

So, the phase transition in one (say, "black") universe happens spontaneously, however due to the
quantum entanglement of two universes ($[\bullet \bullet]$ and $[\circ \circ]$ clusters), the
transition in the second ("white") universe is induced by the vacuum expectation value (vev) of the
macroscopic loop operator. The plateau corresponds to the coexistence of two phases in the 2D
gravity and when the second universe undergoes the induced phase transition, they get separated and
the plateau terminates. Note that we consider the Euclidean version of the 2D gravity, when the
baby universes can be produced.

If we would deal with fluctuating Ising spins coupled to 2D gravity, the results could be borrowed
from the literature. The case when spins live on faces, has been solved in \cite{kazakov}, while
the dual Ising model for spins living on vertices, was treated in \cite{kostov}. In the last case
the model has been reduced to the $O(1)$ loop model. In our system we have spins on the vertices,
hence the dual Ising model \cite{kostov} is more relevant. It admits the two-matrix representation
\cite{kostov}
\beq
Z=\int dXdY e^{{\rm Tr}\,X^2 +\mu {\rm Tr}\,X^3 + {\rm Tr}\,Y^2 +q\,{\rm Tr}\,XY^2}
\label{kostov}
\eeq
where $X$ is the adjacency matrix both for black-black and white-white links, while $Y$ is the
matrix of black-white connections. Besides, we have two additional ingredients compared to
\cite{kostov}: the spins are quenched, and the graph vertex degree is conserved. The degree
conservation can introduced into the matrix model by the linear term with the Lagrangian multiplier
${\rm Tr}\, M X$, playing a role of an external magnetic field similar to \cite{kostov2}. At $Y=0$
we are back to the pure matrix model describing the 2D gravity with the cosmological constant
$\mu$.

The plateau formation has an analogy with the chiral symmetry breaking in QCD with non-vanishing
baryonic chemical potential, $\mu$. Remind few key QCD properties relevant to our system. At $T=0$
the QCD is in the confinement phase with chiral condensate, $\la \bar{\Psi}\Psi \ra \neq 0$. The
popular model of the QCD vacuum is the instanton--antiinstanton ($I\bar{I}$) liquid, represented by
an ensemble of interacting point-like objects of two types. The value of chiral condensate is
presumably expressed in terms of number of connected instanton--antiinstanton pairs \cite{shuryak},
which is the number of black-white links in the network language. The number of fermionic zero
modes at the instanton is an unchanged topological number, meaning fixed vertex degree in network
terms. In the $I\bar{I}$-ensemble zero modes at individual instantons get collectivized. The phase
diagram in QCD is quite rich in the "temperature--chemical potential" plane. At small $\mu$ the
system is in the confinement phase with broken chiral symmetry. By increasing $\mu$, one reaches
mixed phase with two coexisting condensates, $\la \bar{\Psi}\Psi \ra$ and $\la\Psi \Psi\ra$. Such a
mixed phase exists in some region in $\mu$, in which the string tension does not depend on the
density, being the analogue of the network plateau. Further growth of $\mu$ forces the system to
fall down into the so-called  color superconductor phase, with chiral symmetry restoring and QCD
string disappearing. In the QCD color superconductor phase, which is believed to occur in neutron
stars, condensation of quark-quark state happens similar to the Cooper pairing in conventional
superconductors. The phenomenon of chiral symmetry restoration at large $\mu$ apparently can be
described by the effective random matrix model (see \cite{wv} for review). We conjecture that the
two-color network picture could be regarded as a toy model for zero-mode collectivization and can
be related with holographic representation of the chiral symmetry breaking proposed in
\cite{kiritsis}.

\section{Discussion}

In this paper we have described a novel critical phenomenon in statistics of two-color random
network -- the emergence of a wide plateau in the concentration of bonds linking black and white
nodes. The key condition for the plateau development is the graph vertex degree conservation, the
very natural condition for any topological ensemble. Physics behind the phenomenon can be thought
of as the particular mechanism of bonds collectivization in Levin-Wen-type topological
Hamiltonians. It has evident parallels with the effect of fermionic zero modes collectivization,
emerging from localized solutions of Dirac equation in the instanton--antiinstanton QCD ensemble.
Remind, that this effect results in chiral symmetry breaking at small condensate density and its
restoration at large density. It would be highly desirable to find an appropriate matrix model
description of phenomenon, similar to the QCD chiral matrix model describing the spectrum of the
Dirac operator in the $I\bar{I}$ ensemble.

One could consider ensembles of topological defects of different nature like instantons, monopoles
or vertices. The effect of the plateau formation due to the zero modes collectivization is expected
to be universal for them. Recently, formation of the condensate in the "colorless" instanton
ensemble without antiinstantons was investigated \cite{knots}, where it has been found that
microscopic description of condensate involves refined knot invariants. The critical behavior in
such "colorless" topological ensembles inspires the conjecture that the knot recognition by the
topological invariant is different in different phases. It would be very interesting to understand
whether the relation with the knot invariants can be found for colored topological ensembles.

In this work we follow the probe approach for the string, however one could question whether the
critical plateau formation would have stringy meaning if considering the colored network as a
quantum gravity model. The idea to use colored network in this way has been recently suggested in
\cite{trug}. However our model differs from the one proposed in \cite{trug} in two crucial
respects: i) we consider conserved vertex degree, and ii) in our case, contrary to \cite{trug}, the
flow increases the number of triads, since the signs of cosmological constants in our model and in
\cite{trug} are different.

The novel criticality found in colored networks seems to be quite general phenomenon and could have
the practical applications in real life networks where there are sets of separated communities
(black and white nodes). The criticality takes place even in highly asymmetric black-white networks
(i.e. for the parameter $c$ lying in the region about $[0.1, 0.9]$). As an immediate example we
could mention social networks. In this framework, the phenomenon of the plateau formation sounds as
follows. Consider two social communities, and assume that each community desires to increase the
number of triadic "own connections" (simplest "own cliques") measured with some weight (chemical
potential of triads). We predict in such networks existence of the "stability plateau" (i.e.
presence of unavoidable communication between communities), at which the number of links between
different communities is insensitive to the weight of "own connections".
\begin{acknowledgments}

We are grateful to D. Kharzeev, I. Kostov, A. Lokhov, A. Mironov and M. Tamm for fruitful
discussions. S.N was partially supported by the IRSES DIONICOS grant. O.V. and V.A. acknowledge the
support of the Higher School of Economics program for Basic Research. The work of A.G. was
supported in part by grants RBBR-15-02-02092 and Russian Science Foundation grant for IITP
14-050-00150. A.G. thanks the Simons Center for Geometry and Physics for hospitality and support
during the program "Knot homologies, BPS states and Supersymmetric Gauge Theories".

\end{acknowledgments}

\end{document}